\documentstyle[preprint,prl,aps]{revtex}

\draft
\newcommand{\be}{\begin{equation}}
\newcommand{\ee}{\end{equation}}
\newcommand{\ben}{\begin{eqnarray}}
\newcommand{\een}{\end{eqnarray}}

\newcommand{\iii}{\'{\i}}

\begin{document}
\draft
\title{On the Correlations Between Quantum Entanglement and $q$-Information Measures}
\author{J. Batle$^1$, M. Casas$^1$, A. Plastino$^{2}$, and A. R. Plastino$^{1,\,3}$}

\address {
$^1$Departament de F\iii sica, Universitat de les Illes Balears, 07071 Palma de
Mallorca, Spain \\
$^2$Department of Physics, National University La Plata,
  C.C. 727, 1900 La Plata, Argentina  \\
 $^3$Faculty of Astronomy and Geophysics, National University
La Plata and CONICET,  C.C. 727, 1900 La Plata}


\maketitle

\begin{abstract}
In the present study we revisit the application of the
$q$-information measures $R_q$ of  R\'enyi's and $S_q$ of Tsallis'
to the discussion of special features of two qubits systems.  More
specifically, we study the correlations between the
$q$-information measures and the entanglement of formation of a
general (pure or mixed) state $\rho$ describing a system of two
qubits. The analysis  uses a Monte Carlo procedure involving the
15-dimensional 2-qubits space of pure and mixed states, under the
assumption that these states are uniformly distributed according
to the product measure recently introduced by Zyczkowski {\it et
al} [Phys. Rev. A {\bf 58} (1998) 883]. \vskip 2mm {\bf Pacs:}
03.67.-a; 89.70.+c; 03.65.Bz

\end{abstract}
\newpage
Important tools have been developed in recent years for the systematic
exploration of the entanglement properties of composite quantum systems
\cite{ZHS98,Z99}. Quantum entanglement is a physical resource associated with
the peculiar nonclassical correlations
  that may exist between separated quantum systems
\cite{GD02,NC00}. Entanglement is indeed the basic resource
required to implement quantum information processes
\cite{GD02,NC00,G99,LPS98,W98,BDMT98,WC97}. A state of a composite
quantum system is called ``entangled" if it can not be represented
as a mixture of factorizable pure states. Otherwise, the state is
called separable. The above definition is physically meaningful
because entangled states (unlike separable states) cannot be
prepared locally by acting on each subsystem individually. In
particular, for bipartite pure states $\vert \Psi \rangle_{AB}$
one finds that they are entangled if their Schmidt number is
greater that one. Otherwise, they are separable and their
associated reduced density matrices $\hat \rho_A$, $\hat \rho_B$
are projectors. Any bipartite pure  state that cannot be expressed
as the direct product $\vert \Psi \rangle_{AB}=\vert \phi
\rangle_A\,\vert \chi \rangle_B$ is entangled and $\hat \rho_A$,
$\hat \rho_B$ represent mixed states.  Two-qubits systems are the
simplest quantum mechanical systems exhibiting the entanglement
phenomenon and play a fundamental role in quantum information
theory. They also provide useful limit cases for testing the
behaviour of more involved systems \cite{TPA02}. The concomitant
space ${\it S}$ of {\it mixed states} is 15-dimensional. Its
entanglement properties being far from trivial, their complete
characterization constitutes a currently active field of research
\cite{ZHS98,Z99,IH00,MJWK01,AS03}.

  Considerable attention has been paid in recent years to the
  application of $q$-entropies to the study of quantum entanglement
  \cite{TPA02,HHH96,HH96,CA97,V99,TLP01,TLB01,AT02,T02,A02,VW02,GG01,CR02}.
  These entropic measures
  incorporate both R\'enyi's \cite{BS93} and Tsallis' \cite{T88,LV98,LSP01}
  families of information measures as special instances (both admitting, in turn,
  Shannon's measure as the particular case associated with the limit
  $q\rightarrow 1$). The early motivation for these studies was
  the development, on the basis of the conditional $q$-entropies,
  of practical separability criteria for density matrices.
  The discovery by Peres of the partial transpose criteria, which for
  two-qubits and qubit-qutrit systems turned out to be both necessary
  and sufficient, rendered that original motivation a bit weaker,
  once it was realized that it is not possible to find a necessary and sufficient
  separability criterium on the basis of just the eigenvalue spectra of
  the three density matrices
  $\hat \rho_{AB}, \,\,\hat \rho_A=Tr_B[\hat \rho_{AB}]$,
  and $\hat \rho_B=Tr_A[\hat \rho_{AB}]$
  associated with a composite system $A\otimes B$ \cite{NK01}. Is is clear,
  however, from the studies reported in
  \cite{TPA02,HHH96,HH96,CA97,V99,TLP01,TLB01,AT02,T02,A02,VW02,GG01,CR02},
  that $q$-entropies do
  play a significant role in the investigation of entanglement phenomena.
  It is our intention in this Communication to investigate the
  degree of correlation between (i) the amount of entanglement $E[\rho_{AB}]$
  exhibited by a two-qubits state $\rho_{AB}$, and (ii)
  the $q$-entropies (or $q$-information measures) of $\rho_{AB}$ (notice that we refer here
  to the total $q$-entropy of the density matrix $\rho_{AB}$ describing
  the composite system as a whole. We shall not consider conditional
  $q$-entropies). It is well known that the amount entanglement and the
  degree mixture (as measured by the $q$-entropies) of a state $\rho_{AB}$
  are independent quantities. However, there is a certain degree of
  correlation among them. States with an increasing degree of mixture tend to
  be less entangled. In point of fact, all two-qubits states with a large enough
  degree of mixture are separable. We want to explore to what extent does
  the strength of the alluded to correlation depend upon the parameter $q$
  characterizing the $q$-entropy used to measure the degree of mixture.
  In particular, we want to find out if there is a special value of $q$
  yielding a better entropy-entanglement correlation than the entropy-entanglement
  correlations associated with other values of $q$. To such an
  end, and for the sake of completeness, a few words regarding
  measures of entanglement are necessary.

  A physically motivated measure of entanglement is
 provided by the entanglement of formation $E[\hat \rho]$  \cite{BDSW96}.
 This measure quantifies the resources needed to create a given
 entangled state $\hat \rho$. That is,
 $E[\hat \rho]$ is equal to the asymptotic limit (for large $n$) of the
 quotient $m/n$, where $m$ is the number of singlet states needed to create $n$
 copies of the state $\hat \rho$ when the optimum procedure based on local
 operations is employed.
 For the particular case of two-qubits states Wootters obtained an explicit
 expression for  $E[\hat \rho]$ in terms of the density matrix
 $\hat \rho$ \cite{WO98}. Wootters' formula reads \cite{WO98}

\be
E[\hat \rho] \, = \, h\left( \frac{1+\sqrt{1-C^2}}{2}\right), \ee

\noindent where $h(x) \, = \, -x \log_2 x \, - \, (1-x)\log_2(1-x)$, and $C$
stands for the {\it concurrence}  of the two-qubits state $\hat \rho$. The
concurrence is given by $C \, = \,
max(0,\lambda_1-\lambda_2-\lambda_3-\lambda_4)$, \,\,$\lambda_i, \,\,\, (i=1,
\ldots 4)$ being the square roots, in decreasing order, of the eigenvalues of
the matrix $\hat \rho \tilde \rho$, with $\tilde \rho \, = \, (\sigma_y \otimes
\sigma_y) \rho^{*} (\sigma_y \otimes \sigma_y)$. The above expression has to be
evaluated by recourse to the matrix elements of $\hat \rho$ computed with
respect to the product basis.

Our investigations will be based upon a Monte Carlo exploration of  ${\cal S}$:
the set of {\it all states, pure and mixed} of a two-qubits
 system. To do this we need to define a proper measure on ${\cal S}$. The space
of all (pure and mixed) states $\rho$ of a quantum system described by an
$N$-dimensional Hilbert space can be regarded as a product space ${\cal S} =
{\cal P} \times \Delta$ \cite{ZHS98,Z99,BCPP02a,BCPP02b}. Here $\cal P$ stands
for the family of all complete sets of orthonormal projectors $\{ \hat
P_i\}_{i=1}^N$, $\sum_i \hat P_i = I$ ($I$ being the identity matrix). $\Delta$
is the set of all real $N$-uples $\{\lambda_1, \ldots, \lambda_N \}$, with $0
\le \lambda_i \le 1$, and $\sum_i \lambda_i = 1$. The general state in ${\cal
S}$ is of the form $\rho =\sum_i \lambda_i P_i$.  The Haar measure on the group
of unitary matrices $U(N)$ induces a unique, uniform measure $\nu$ on the set
${\cal P}$ \cite{PZK98}. On the other hand, since the simplex $\Delta $ is a
subset of a $(N-1)$-dimensional hyperplane of ${\cal R}^N$, the standard
normalized Lebesgue measure ${\cal L}_{N-1}$ on ${\cal R}^{N-1}$ provides a
natural measure for $\Delta$. The aforementioned measures on $\cal P$ and
$\Delta$ lead to a natural measure,

\be \label{mumesu} \mu= \nu \times {\cal L}_{N-1}, \ee

\noindent
on the set $\cal S$ of quantum states \cite{ZHS98,Z99,slater}.

  In the present investigation we deal with the case $N=4$.
  Our present considerations are based on the assumption
 that the uniform distribution of states of a two-qubit system
 is the one determined by the measure $\mu$. Thus, in our
 numerical computations we are going to randomly generate
 states of a two-qubits system according to the measure
 $\mu$ and  study the relation between the entanglement properties of
 these states, on the one hand,
and
\begin{itemize}
\item 1) the Tsallis $q$-entropy  \be \label{tsallis}
  S_q \, = \, \frac{1}{q-1}\bigl(1-\omega_q \bigr),
\,\, $with$ \,\,
  \omega_q \, = \, Tr \left( \hat \rho^q \right),
  \ee
  and
  \item 2) the R\'enyi $q$-entropy
\be R_q= \frac{1}{1-q}\,\ln{[1+(1-q)\,S_q]},\ee
on the other one.
 \end{itemize}

  Most recent research efforts dealing with the relationship between the degree
  of mixture and the amount of entanglement focus on the behaviour, as a
  function of the degree of mixture, of the entanglement properties exhibited
  by the set of states endowed with a given amount of mixedness. For instance,
  they consider the behaviour, as a function of the degree of mixture
  (as measured, for instance, by $S_2$), of the average entanglement of
  those states characterized by a given value of $S_2$.
  Here we are going to adopt, in a sense, the reciprocal (and complementary)
  point of view. We are going to study the behaviour, as a function of $C^2$,
  of the entropic properties associated with the set of states characterized
  by a given value of $C^2$. This vantage point will enable us to clarify some
  aspects of the $q$-dependence of the entanglement-mixedness correlation. In
  particular, we want to asses, for different $q$-values, how sensitive are
  the average entropic properties to the value of the entanglement of
  formation (or, equivalently, to the value of the squared concurrence $C^2$).

  First of all, we have randomly generated a large number of states
(according to the measure $\mu $ given by expression (\ref{mumesu})) and
plotted them, for several $q$-values, in the $(C^2,R_q)$-plane. Each point in
Fig. 1 correspond to a particular state in the state space ${\cal S}$. The
outcome is a series of ``bands", one for each $q$. The top band corresponds to
$q=0.5$. At the bottom we find that for $q=10$. The remaining ones arrange
themselves in between, in monotonic fashion.  One appreciates the fact that,
for the different $q$-values, we find, given a certain amount of entanglement
as measured by $C^2$, states of larger and larger entropies $R_q$ as $q$
diminishes. From the point of view of information theory this would entail that
information about our states is lost as $q$ decreases. If we now consider
averages over all states that share a given concurrence we are led to consider
Fig. 2.

   We computed, as a function of $C^2$, the average value of
the R\'enyi entropy $R_q$ associated with the set of states
endowed with a given value of the squared concurrence $C^2$. The
results are exhibited in Fig. 2 (solid lines), where the mean
value $\langle R_q \rangle$ is plotted against $C^2$, for $q
=0.5,\,1,\,2,\,10,$ and $\infty$. As stated, the averages are
taken over all the states $\hat \rho \in {\cal S}$ that are
characterized by a fixed concurrence-value. For all $q$ the
average entropies diminish as $C$ grows. This behaviour is
consistent with the fact that states of increasing entropy tend to
exhibit a decreasing amount of entanglement
\cite{ZHS98,BCPP02a,BCPP02b}. As $q$ grows, the average entropy
decreases, for any $C^2$, although the decreasing tendency slows
down for large $q$-values.
 Many recent efforts dealing with the relationship between $q$-entropies and
 entanglement were restricted to states  $\rho_{\rm Bell}$ diagonal in the
 Bell basis. For such states, both the $R_q$ entropy and the squared
concurrence $C^2$ depend solely upon $\rho_{\rm Bell}$'s largest
eigenvalue, so that $R_q$ can be expressed as a function of $C^2$.
The dashed line in Fig. 2 depicts the functional dependence of the
$R_{\infty}$ R\'enyi entropy, as a function of $C^2$, for
two-qubits states diagonal in the Bell basis. It is instructive to
compare, in Fig. 2, the curve corresponding to states diagonal in
the Bell basis with the curve corresponding (with $q=\infty $) to
all two-qubit states. It can can be appreciated that these two
curves, even if sharing the same qualitative appearance, differ to
a considerable extent.

 For the sake of comparison, we plotted in Fig. 3 the mean value
 $\langle S_q \rangle$ of Tsallis' entropy, as a function of $C^2$, for
 $q =0.5,\,1,\,2,$ and $10$. Again, for each value of $C^2$, the entropy's
 average was computed over all those states characterized by
 that particular $C^2$-value. Notice that for large $q$-values, the Tsallis
entropy is approximately constant for all $C^2$ values, while the
R\'enyi one seems to be much more sensitive in this respect.
Entropies tend to vanish for $C^2 \to 1$, because only pure states
can reach the maximum concurrence value. In the inset of Fig. 3 we
depict the behaviour of $\langle S_q \rangle_{C^2}$ as a function
of $1/q$ for a given value of the concurrence ($C^2 =0.6$), thus
illustrating the fact that the mean entropy is a monotonically
decreasing function of $q$. For large $q$-values the Tsallis
entropy cannot discriminate between different degrees of
entanglement for states with $C^2 <1$, while R\'enyi's measure can
do it. This fact is related to an important difference between the
behaviours, as a function of the parameter $q$, of R\'enyi's $R_q$
and Tsallis' $S_q$ entropies. The maximum value $R_q^{\rm max}$
attainable by R\'enyi's entropy (corresponding to the
equi-probability distribution) is independent of $q$,

\be
R^{\rm max}_q \, = \,  -\ln N, \ee

\noindent where $N$ is the total number of accesible states. On the contrary,
the maximum value reachable by $S_q$ does depend upon $q$,

\be
S^{\rm max}_q \, = \, \frac{1 - N^{1-q}}{ (q-1)}.
 \ee

\noindent Clearly, $S^{\rm max}_q \rightarrow 0$ for $q \rightarrow \infty$.
One may think that the $q$-dependence of $S^{\rm max}_q$ may be appropriately
taken into account if one considers (instead of Tsallis' entropy itself), a
{\it normalized} Tsallis' entropy (see Fig. 4),

\be \label{normal} S'_q = \frac{S_q}{S_q^{max}}, \ee

\noindent  For instance, in the case of two qubits one has,

\be \label{2q} S_q^{max} = \frac{1 - 4^{1-q}}{ (q-1)}, \ee

\noindent and we deal then with \be \label{normal1} S'_q = \frac{1-Tr [\hat
\rho^q]}{1-4^{1-q}} = \frac{1-\{[Tr (\hat \rho^q)]^{1/q}\}^q}{1-4^{1-q}}. \ee
Consider now the limit $q \rightarrow \infty$ for a density matrix $\hat \rho$
corresponding to a state of fixed concurrence $C$. In such a process one
immediately appreciates the fact that $[Tr (\hat \rho^q)]^{1/q} \rightarrow
\lambda_{max}$, where $\lambda_{max}$ is the largest eigenvalue of $\hat
\rho^q$. Thus, the limiting value we reach is \be \label{limite} S'_q
\rightarrow [1 - (\lambda_{max})^q], \ee and we see that this is always equal
to unity for all $C^2<1$ and vanishes exactly if $C^2=1$ (see Fig. 4).
Consequently, even employing the normalized $S'_q$, the information concerning
the entropy-entanglement correlation tends to disappear in the $q\rightarrow
\infty$ limit.

 We see that, with regards to the analysis of the entropy-entanglement
correlations of quantum bipartite systems (remember that here we
are talking about the total entropy of the whole composite
system), Tsallis' and R\'enyi's entropies behave in different
ways. It is instructive to compare this behaviour of the
$q$-entropies, to their behaviour with regards to the relationship
between separability and conditional $q$-entropies. What matters
in this last case {\it is the sign of the conditional
$q$-entropies}. If a classical composite system $A \times B$ is
described by a suitable probability distribution $\{ p_{AB} \}$
(defined over the concomitant phase space), the $q$-information
measure (or $q$-entropy) associated to the whole system is always
equal or greater than that pertaining to its subsystems, i.e.,
those associated to $\{ p_A \}$ or $\{ p_B \}$. In other words,
the conditional entropy

 \be \label{qonditional}
 S_q[A|B] \, = \, S_q[p_{AB}] \, - \, S_q[p_{B}],
 \ee

 \noindent
 is always positive. This is also the case for {\it separable} states
 $\hat \rho_{AB}$ of a composite quantum system $A\otimes B$ \cite{NK01,VW02}.
 In contrast, a subsystem of a quantum system described by an entangled
 state may have an entropy greater than the entropy of the whole system. That
 is to say, {\it entangled states may have negative conditional $q$-entropies}.
 Now, since R\'enyi's entropy is a monotonically increasing function of Tsallis'
 measure, it is clear that the sign of the conditional R\'enyi's entropy
 associated with a quantum state $\hat \rho_{AB}$,
 $R_q[A|B]=R_q[\hat \rho_{AB}]-R_q[\hat \rho_{B}]$,
 will be the same as the sign of the conditional Tsallis' entropy,
 $S_q[A|B]=S_q[\hat \rho_{AB}]-S_q[\hat \rho_{B}]$. Hence, as far as the
 relationship between conditional entropies and separability
 (which depends only upon the sign of the conditional $q$-entropies)
 is concerned, Tsallis' and R\'enyi's measures are equivalent.

  Returning to our discussion of the connection between entanglement and
(total) $q$-entropies of bipartite quantum systems, we have seen
that R\'enyi's entropy is particularly well suited for (i)
discussing the $q\rightarrow \infty$ limit and (ii) studying the
$q$-dependence of the entropy-entanglement correlations. For these
reasons, in the rest of the present contribution we will focus
upon R\'enyi entropy.

 We tackle now the question of the dispersion around these entropic averages.
Fig. 5  is a graph of the dispersions

\be \label{dispersion} \sigma^{(R)}_q \, = \,
 \left[\langle R_q^2 \rangle-
\langle R_q \rangle^2 \right]^{1/2}, \ee

\noindent as a function of $C^2$, for the same $q$-values of Fig. 2. We see
that the size of the dispersions diminishes rather rapidly as $C^2$ increases
towards unity. Also, dispersions tend to become smaller as $q$ grows. This
suggests that, as $q$ increases, the correlation between $\langle R_q \rangle$
and entanglement improves. A similar tendency, but in the case of $S_q$, was
detected in \cite{CR02}.

  In order to estimate in a quantitative the sensitiveness of the average
$q$-entropy to changes in the value of $C^2$, we computed the
derivatives with respect to $C^2$  of the average value of
R\'enyi's entropy associated with states of given $C^2$,

  \be \label{derivadas} \frac{d\langle R_q \rangle}{d(C^2)} . \ee

\noindent In Fig. 6 we plot the above derivatives, against $C^2$, for $q=0.5,
1, 2, 10$, and $\infty$. These derivatives fall abruptly to zero, in the
vicinity of the origin, as $C^2$ diminishes. As a counterpart, for all $q$, the
derivatives exhibit a rapid growth  with $C^2$ for small values of the
concurrence. This growing tendency stabilizes itself and, for $q$ large enough,
saturation is reached.

Now let us assume that we know the value of the entropy $R_q[\rho]$ of certain
state $\rho$. How useful is this knowledge in order to infer the value of
$C^2$?. In other words, how good is $R_q$ as an ``indicator" of entanglement?
It has been suggested that $q=\infty $ provides a better ``indicator" of
entanglement than other values of $q$ \cite{GG01,CR02}. There are two
ingredients that must be taken into account in order to determine the $q$-value
yielding the best entropic ``indicator" of entanglement. On the one hand, the
sensitivity of the entropic mean value $\langle R_q \rangle$ to increments in
$C^2$, as measured by the derivative $d\langle R_q\rangle/d(C^2)$. On the other
hand, the dispersion $\sigma^{(R)}_q$, given by (\ref{dispersion}). A given
$q$-value would lead to a good entropic ``indicator" if it corresponds to (i) a
large value of $d\langle R_q\rangle/d(C^2)$, and (ii) a small value of
$\sigma^{(R)}_q$.  These two factors are appropriately taken into account if we
compute the ratio

\be \label{ratio} r \, = \, \left|\frac{\sigma^{(R)}_q}{d\langle R_q
\rangle/d(C^2)}\right|, \ee

\noindent
 between the dispersions depicted in Fig. 5 and the derivative of Fig. 6.
 The ratio $r$ provides a quantitative measure for the
strength of the entropic-entanglement correlations. The quantity $r$
constitutes an estimate of the smallest increment $\Delta C^2$ in the squared
concurrence which is associated with an appreciable change in $R_q$. In order
to clarify this last assertion, an analogy with the uncertainty associated with
the measurement of time in quantum mechanics can be established. Let us assume
that we can measure an observable $\hat A$. Then, the time uncertainty $\Delta
t$ depends upon two quantities, (i) the time derivative of the expectation
value of the observable, $d \langle \hat A \rangle/dt$, and (ii) the
uncertainty of the observable, $\Delta \hat A = [\langle \hat A^2 \rangle
-\langle \hat A \rangle^2]^{1/2}$. The time uncertainty is given by \cite{M61}

\be \label{deltime} \Delta t \, = \, \frac{\Delta \hat A}{d \langle \hat A
\rangle/dt}
 \ee

\noindent The above expression for $\Delta t$ gives an estimation
of the smallest time interval that can be detected from
measurements of the observable $\hat A$. In the analogy we want to
establish, $C^2$ plays the role of $t$, and $R_q$ plays the role
of the observable $A$. The ratio $r$ is depicted in Fig. 7, as a
function of $C^2$, for $q=1$ and $q=\infty$. The two upper curves
in Fig. 7 correspond to the $r$-values obtained when all the
states in the two-qubits state-space ${\cal S}$ are considered. On
the other hand, the lower curves are the ones obtained when the
computation of $r$ is restricted to states diagonal in the Bell
basis. When all states in ${\cal S}$ are considered, the values of
$r$ associated with $q=\infty$ are seen to be smaller than the
values corresponding to $q=1$, which can be construed as meaning
that the $q$-entropies with $q=\infty$ can indeed be regarded as
better ``indicators" of entanglement than the $q$-entropies
associated with finite values of $q$, as was previously suggested
in \cite{GG01,CR02}. Alas, the results depicted in Fig. 7 indicate
that this improvement of the entropy-entanglement correlation
associated with $q=\infty$ is not considerable. The usefulness of
$q$-entropies with $(q\rightarrow \infty)$ as an ``indicator" of
entanglement was proposed in \cite{GG01} on the basis of the
behaviour of states diagonal in the Bell basis. As already
mentioned, the squared concurrence $C^2$ of states $\rho_{\rm Bell
}$ diagonal in the Bell basis can be expressed as a function of
$R_{\infty}$, since both these quantities depend solely on the
largest eigenvalue $\lambda_m$ of $\rho_{\rm Bell }$ (in
particular, $R_{\infty}=-\ln \lambda_m$). This means that, as
pointed out in \cite{GG01,CR02}, for states diagonal on the Bell
basis there is a perfect correlation between $C^2$ and
$(q\!=\!\infty)$-entropies (and, consequently, $r$ vanishes). This
implies that, when restricting our considerations {\it only} to
states diagonal in the Bell basis, the entropy-entanglement
correlation is much more strong for $q=\infty$ than for other
values of $q$. States diagonal in the Bell basis are important for
many reasons, but their properties are by no means typical of the
totality of the state-space ${\cal S}$. See for instance, as
depicted in Fig.7, the behavior of $r$ (for $q=1$) associated with
(i) all states in ${\cal S}$ and (ii) states diagonal in the Bell
basis. There are remarkable differences between the two cases.

We  thus find ourselves in a position to assert that the
relationship between the $q$-entropies and the amount of
entanglement exhibited by the family of states diagonal in the
Bell basis does not constitute a reliable guide to infer the
typical behavior of states in the two-qubits state-space ${\cal
S}$. When considering the complete state-space ${\cal S}$, the
${q\!=\!\infty}$-entropies turn out to be only a slightly better,
as entanglement ``indicators", than the entropies associated with
other values of $q$.

\vskip 0.5 cm
  We summarize now our present considerations.
  By recourse to a Monte Carlo procedure we have studied the $q$-dependence of
the correlations exhibited by two-qubits states between (i) the
amount of
 entanglement and (ii) the $q$-entropies. It was previously conjectured by
 other researchers, on the basis of the study of states diagonal in the Bell
 basis, that the $q$-entropies associated with $q=\infty $ are better
 ``indicators" of entanglement than the entropies corresponding to finite
 values of $q$. In other words, it was suggested that the $q$-entropy with
 $q=\infty $ exhibits a stronger correlation with entanglement than the other
 $q$ entropies. By a comprehensive numerical survey of the complete (pure and
 mixed) state-space of two-qubits, we have shown  here that the alluded to
 conjecture is indeed correct. However, when globally considering the whole state-space
 the advantage, as an entanglement indicator, of
$(q\!=\!\infty)$-entropy
 turns out to be much smaller than what can be inferred from the sole study of
 states diagonal in the Bell basis. This constitutes an instructive example of
 the perils that entails trying to infer  typical properties of general two-qubits
 states from the study of just a particular family of states, such as those diagonal
 in the Bell basis.

\acknowledgments This work was partially supported by the MCyT
grant BFM2002-03241 (Spain), and by CONICET (Argentine Agency).
One of us (ARP) thanks financial support from the MECyD grant
SAB2001-0106.


\noindent {\bf FIGURE CAPTIONS}

\vskip 0.5cm

\noindent Fig.1- R\'enyi entropy $R_q$ vs. the squared concurrence
$C^2$ for all two-qubits states and several $q$ values. All
depicted quantities are dimensionless.

\vskip 0.5cm

\noindent Fig.2- Average value of the R\'enyi entropy $\langle R_q
\rangle$ of all states with a given squared concurrence $C^2$, as
a function of $C^2$, and for several $q$-values (solid lines). The
dashed line depicts the functional dependence of the $R_{\infty}$
R\'enyi entropy, as a function of $C^2$, for two-qubits states
diagonal in the Bell basis. All depicted quantities are
dimensionless.

\vskip 0.5cm

\noindent Fig.3- Average value of the Tsallis' entropy $\langle
S_q \rangle$ of all states with a given squared concurrence $C^2$,
as a function of $C^2$, and for several $q$-values. The inset
shows $\langle S_q \rangle$ vs. $1/q$ for the particular value of
the squared concurrence $C^2 = 0.6$. All depicted quantities are
dimensionless.

 \vskip 0.5cm

\noindent Fig.4- Average value of the normalized Tsallis entropy
$\langle S_q \rangle/S_{q{\rm max}}$ vs. $C^2$, for several
$q$-values.  All depicted quantities are dimensionless.

\vskip 0.5cm

\noindent Fig.5- Dispersion of the R\'enyi entropy
$\sigma^{(R)}_q=\left[\langle R_q^2 \rangle- \langle R_q \rangle^2
\right]^{1/2}$ for all qubits states with a given $C^2$, as a
function of $C^2$, and for several $q$-values.  All depicted
quantities are dimensionless.

\vskip 0.5cm

\noindent Fig.6- The derivative $d\langle R_q \rangle/d(C^2)$, as
a function of the squared concurrence $C^2$, for several values of
the $q$-parameter. All depicted quantities are dimensionless.

 \vskip 0.5cm

\noindent Fig.7- The absolute value of the quotient $r =
\left|\frac{\sigma^{(R)}_q}{d\langle R_q \rangle/d(C^2)}\right|$,
as a function of the squared concurrence $C^2$, for $q=1$ and
$q=\infty$. The two upper curves correspond to all states in the
two-qubit state-space ${\cal S}$. The lower curves correspond to
states diagonal in the Bell basis. All depicted quantities are
dimensionless.


\begin{thebibliography}{}


\bibitem{ZHS98} Karol Zyczkowski, P. Horodecki, A. Sanpera, and M. Lewenstein,
Phys. Rev. A {\bf 58}, 883 (1998).

\bibitem{Z99} Karol Zyczkowski, Phys. Rev. A {\bf 60},  3496 (1999).

\bibitem{GD02} A. Galindo and M.A. Martin-Delgado, Rev. Mod. Phys. {\bf 74}, 347
(2002).

\bibitem{NC00} M.A. Nielsen and I.L. Chuang, {\it Quantum Computation
and Quantum Information}, (Cambridge University Press, Cambridge, 2000).

\bibitem{G99} J. Gruska, {\it Quantum Computing}, (McGraw-Hill, London, 1999).

\bibitem{LPS98} Hoi-Kwong Lo, S. Popescu  and  T. Spiller (Editors)
{\it Introduction to Quantum Computation and Information} (World
Scientific, River Edge, 1998).

\bibitem{W98} C.P. Williams (Editor) {\it Quantum Computing and Quantum
Communications} (Springer, Berlin, 1998).

\bibitem{BDMT98}  G.P. Berman, G.D. Doolen, R. Mainieri, and V.I.
Tsifrinovich,  {\it Introduction to Quantum Computers} (World Scientific,
Singapore, 1998).

\bibitem{WC97} C.P. Williams and S.H. Clearwater,  {\it Explorations in
Quantum Computing} (Springer, New  York, 1997).


\bibitem{TPA02} C. Tsallis, D. Prato, and C. Anteneodo, Eur. Phys. J. B {\it
29} (2002) 605.

\bibitem{IH00} S. Ishizaka, T. Hiroshima,
Phys. Rev. A 62 (2000) 022310.


\bibitem{MJWK01} W.J. Munro, D.F.V. James, A.G. White, P.G. Kwiat,
Phys. Rev. A 64 (2001) 030302.

\bibitem{AS03} C. Anteneodo and M.C. Souza, J. Opt. B 5 (2003) 73.


\bibitem{HHH96} R. Horodecki, P. Horodecki, and M. Horodecki,
Phys. Lett. A {\bf 210}, 377 (1996).

\bibitem{HH96} R. Horodecki and M. Horodecki, Phys. Rev. A {\bf 54}, 1838 (1996).

\bibitem{CA97} N. Cerf and C. Adami, Phys. Rev. Lett. {\bf 79}, 5194 (1997).

\bibitem{V99} A. Vidiella-Barranco, Phys. Lett. A {\bf 260}, 335 (1999).

\bibitem{TLP01} C. Tsallis, P.W. Lamberti, D. Prato, Physica A
{\bf 295}, 158 (2001).

\bibitem{TLB01} C. Tsallis, S. Lloyd, and M. Baranger, Phys. Rev.
A {\bf 63}, 042104 (2001).

\bibitem{AT02} F.C. Alcaraz and C. Tsallis, Phys. Lett. A {\it 301} (2002) 105.

\bibitem{T02} B.M. Terhal, Theor. Comp. Sci. {\bf 287}, 313 (2002).

\bibitem{A02} S. Abe, Phys. Rev. A {\bf 65}, 052323 (2002).

\bibitem{VW02} K.G.H. Vollbrecht and M.M. Wolf,
{\it Conditional Entropies and their Relation to Entanglement Criteria}, arXiv:
quant-ph/0202058.

\bibitem{GG01} F. Giraldi and P. Gringolini, Phys. Rev. E {\bf 64}, 2310 (2001).

\bibitem{CR02} N. Canosa and R. Rossignoli, Phys. Rev. Lett. {\bf 88},  170401 (2002).


\bibitem{BS93}  C. Beck and F. Schlogl, {\it Thermodynamics of Chaotic
Systems}, (Cambridge University Press, Cambridge, 1993).

\bibitem{T88} C. Tsallis, J. Stat. Phys. {\bf 52}, 479 (1988).

\bibitem{LV98} P.T. Landsberg and V. Vedral, Phys. Lett. A
{\bf 247}, 211 (1998).

\bibitem{LSP01} J.A.S. Lima, R. Silva, and A.R. Plastino,
Phys. Rev. Lett. {\bf 86}, 2938 (2001).


\bibitem{NK01} M.A. Nielsen and J. Kempe, Phys. Rev. Lett. {\bf 86}, 5184
(2001).


\bibitem{BDSW96}  C.H. Bennett, D.P. DiVicenzo, J. Smolin, and  W.K.
Wootters,  Phys. Rev. A {\bf 54}, 3824 (1996).

\bibitem{WO98} W.K. Wootters, Phys. Rev. Lett. {\bf 80}, 2245 (1998).


\bibitem{BCPP02a} J. Batle, M. Casas, A.R. Plastino, and A. Plastino,
 Phys. Lett. A {\bf 296}, 251 (2002).

\bibitem{BCPP02b} J. Batle, M. Casas, A.R. Plastino, and A. Plastino,
  Phys. Lett. A {\bf 298},  301 (2002).

  \bibitem{PZK98} M. Pozniak, K. Zyczkowski, and M. Kus,
J. Phys A {\bf 31}, 1059 (1998).

 \bibitem{slater} We must mention that Slater has  argued  \cite{slater1,slater2} that, in
 analogy to the classical use of the volume element of the Fisher information
 metric as Jeffreys' prior \cite{jef} in Bayesian theory, a natural measure on
 the quantum states would be the volume element of the Bures metric.
 \cite{bures}.

\bibitem{slater1} P. B. Slater, J. Phys. A {\bf 32}, 5261 (1999).

\bibitem{slater2} P. B. Slater, Eur. Phys. J. B {\bf 17}, 471 (2000).

\bibitem{jef} R. E. Kass, Statist. Sci.  {\bf 4}, 188 (1989).

\bibitem{bures} M. H\"ubner, Phys. Lett. A {\bf 163}, 239 (1992); {\bf 179}, 221 (1993).


\bibitem{M61} A. Messiah, {\it Quantum Mechanics} (North-Holland, Amsterdam,
1961).


\end{thebibliography}
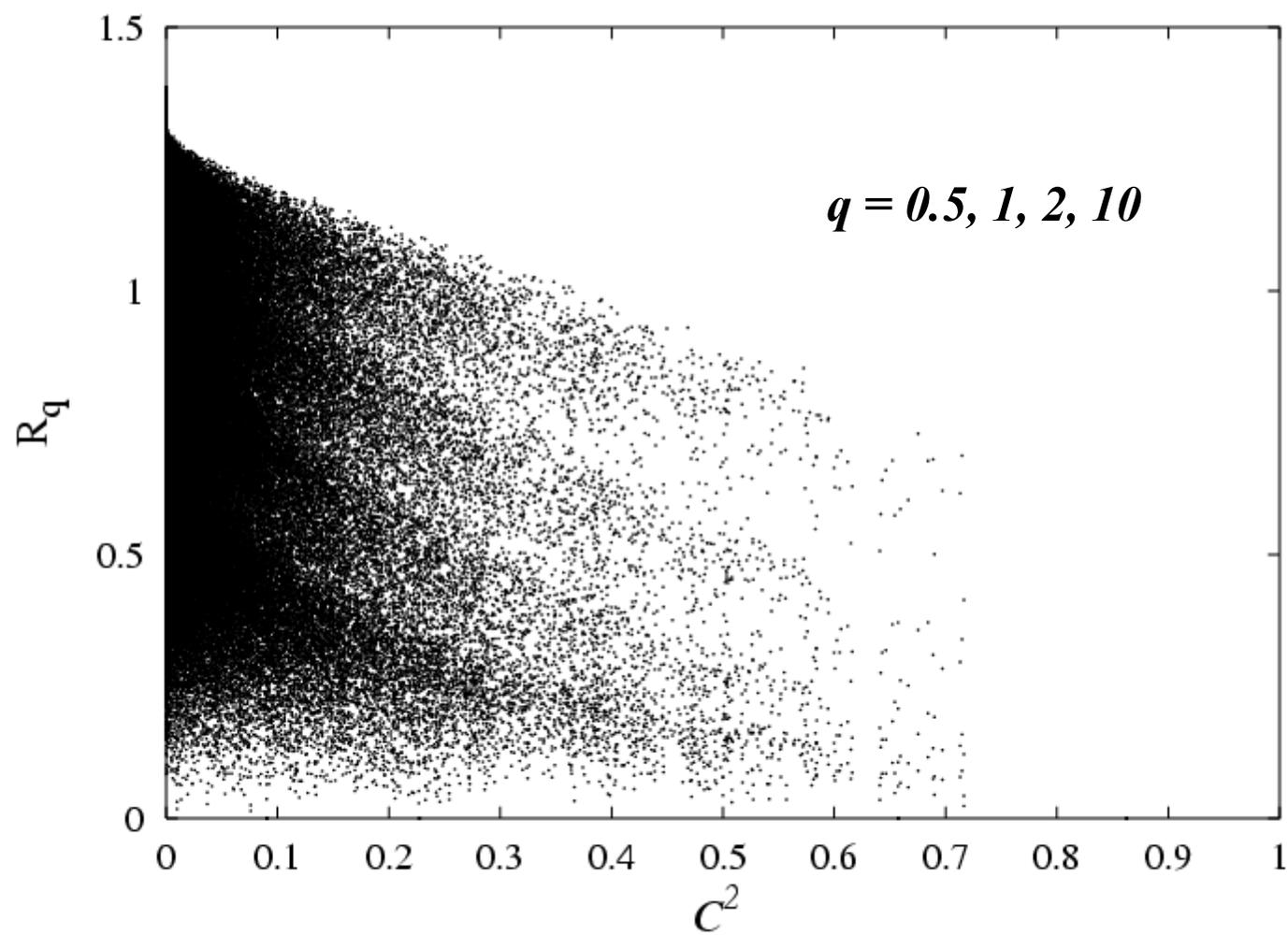
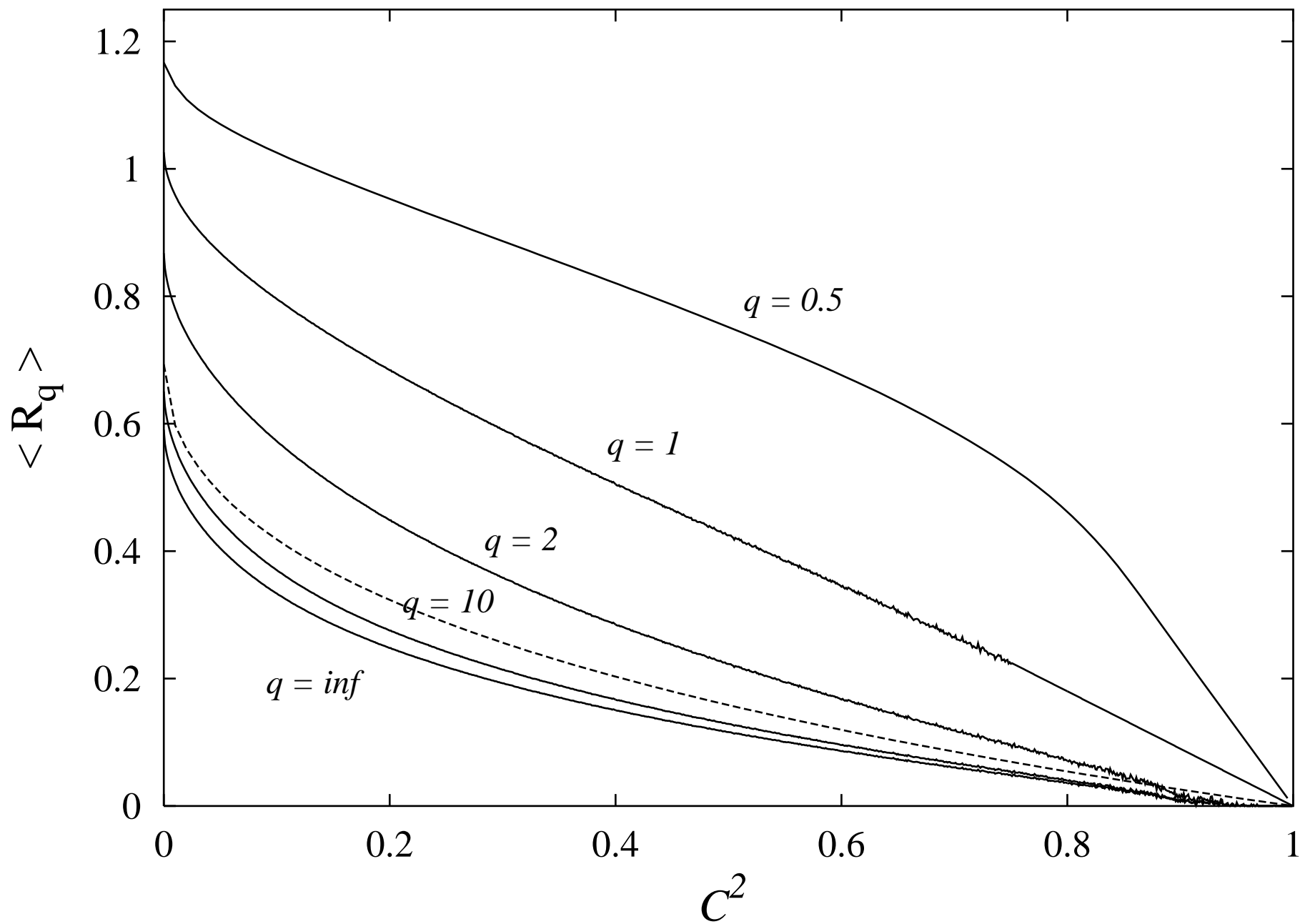
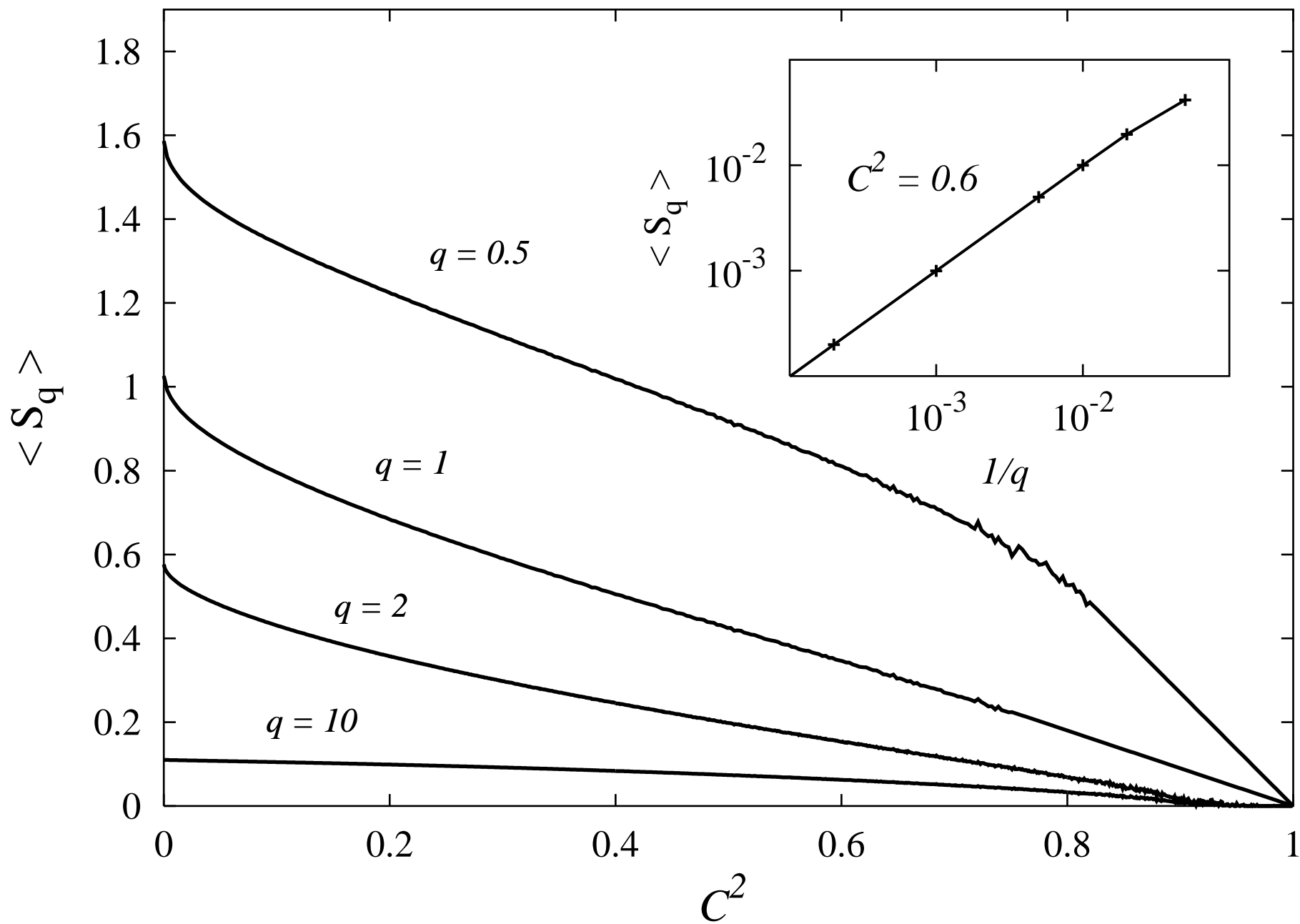
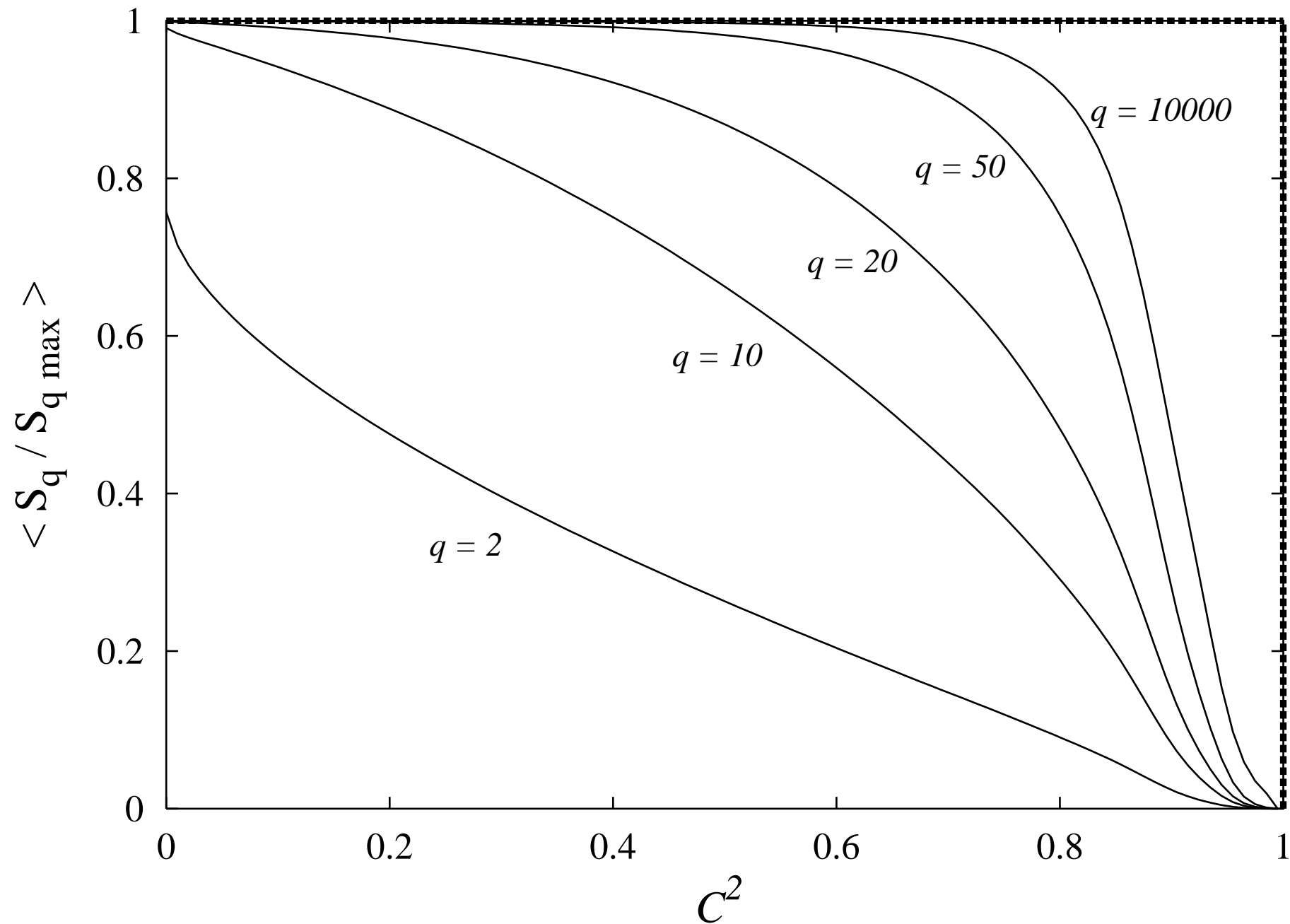
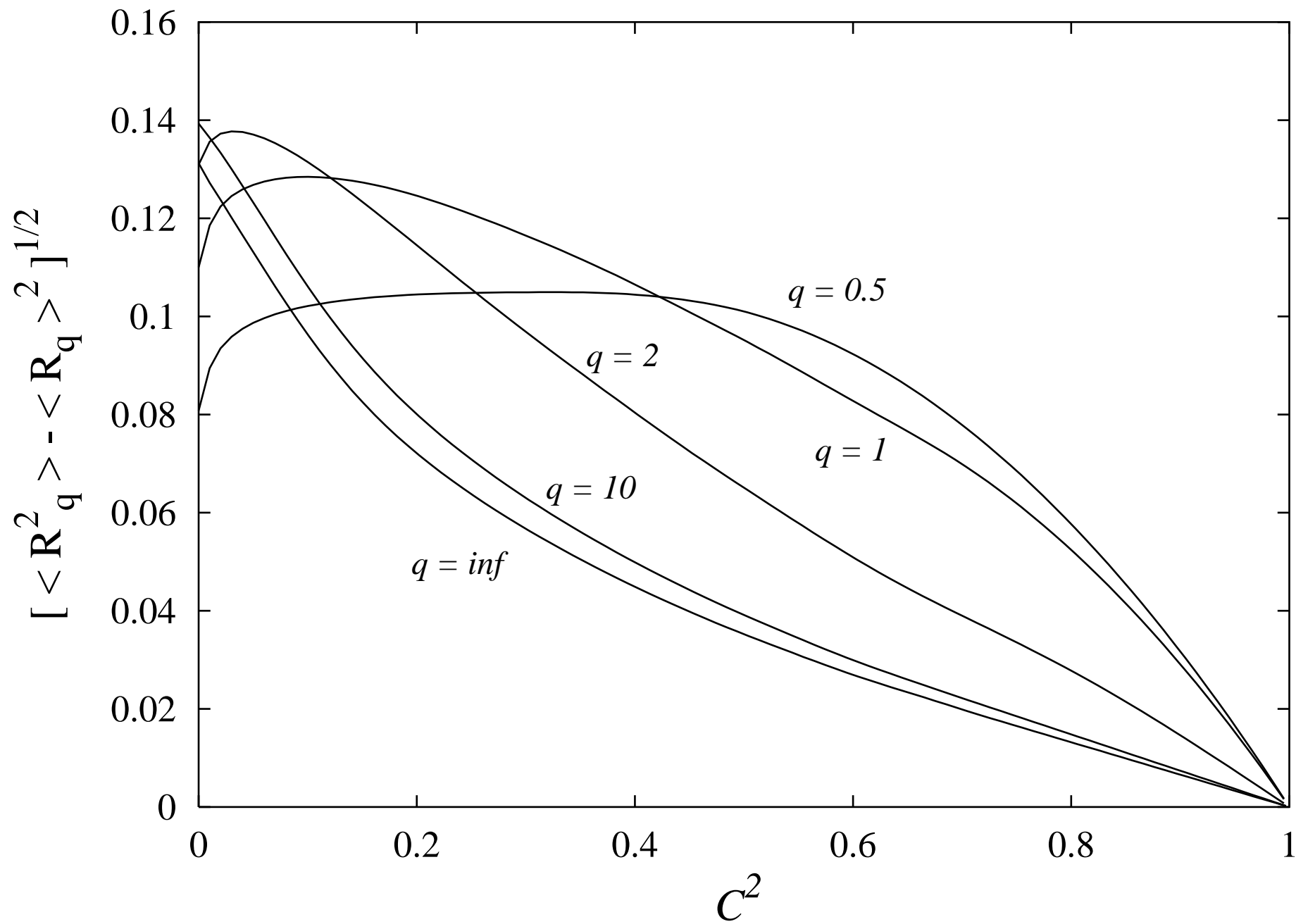
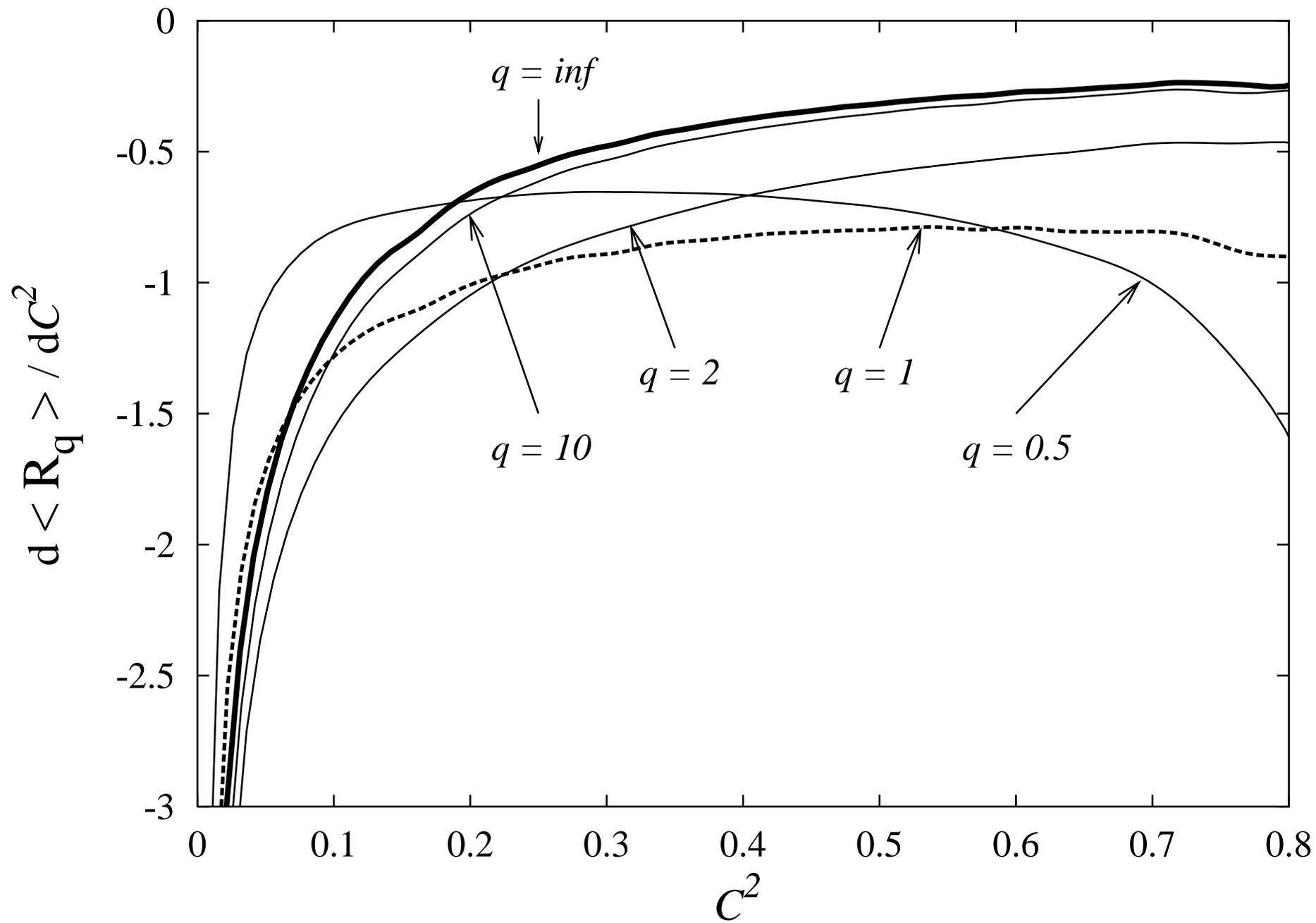
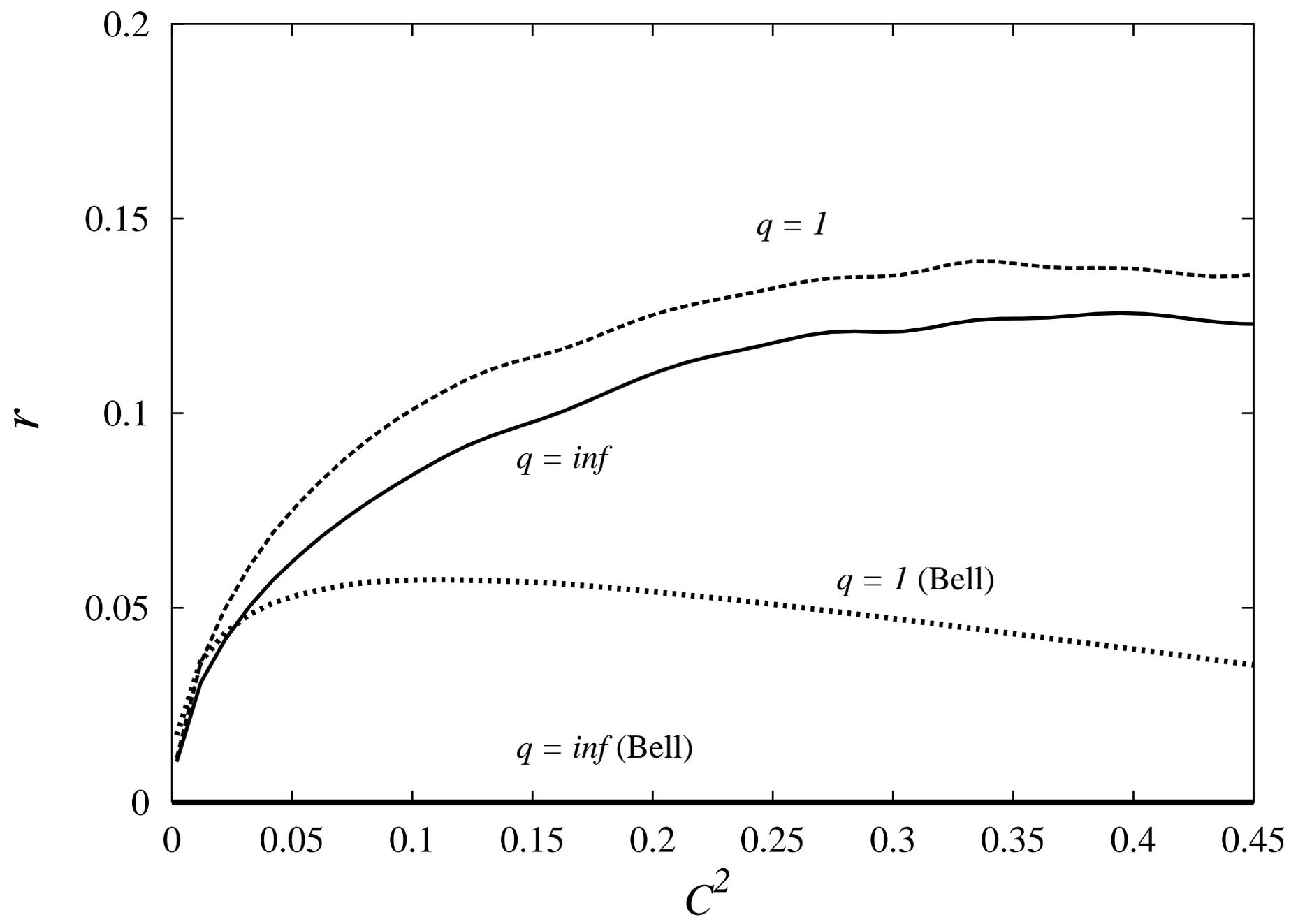
\end{document}